\documentclass[]{spie}  
\usepackage[]{graphicx}
\usepackage{multirow}

\title{Imaging and burst location with the EXIST high-energy telescope  } 

\author{Gerald K. Skinner\supit{a,b}, Scott Barthelmy\supit{c},  Mark H. Finger\supit{d} , Jae Sub Hong\supit{e} , Garrett Jernigan\supit{f} , Stephen~J.~Sturner\supit{a,g} , Branden T.  Allen\supit{e} and Jonathan E.  Grindlay\supit{e}
\skiplinehalf
\supit{a}   CRESST  \&  NASA-GSFC, Greenbelt, MD 20771, USA; \\
\supit{b}  Department of Astronomy, Univ. Md.,  College Park, MD 20742, USA; \\
\supit{c}   NASA-GSFC, Greenbelt, MD 20771, USA; \\
\supit{d}   USRA \& NASA-MSFC, Huntsville, AL 35812, USA; \\
\supit{e}  Harvard-Smithsonian Center for Astrophysics, 60 Garden St., Cambridge, MA 02138 USA; \\
\supit{f}  Space Science Lab., UCB, Berkeley, CA 94720, USA;\\
\supit{g}  Univ. Md. Baltimore County,  USA
}

\authorinfo{Send correspondence to G.K.S:  E-mail: skinner@milkyway.gsfc.nasa.gov }

\newcommand{\degr} {$^\circ$}

\newcommand{\bird}{1996A&AS..117..131B}
\newcommand{\fenimore}{1987ApOpt..26.2760F}
\newcommand{\kutyrev}{kutyrevspie}
\newcommand{\skinnerao}{2008ApOpt..47.2739S}
\newcommand{\skinnercapri}{1995ExA.....6....1S}
\newcommand{\skinnergrindlay}{1993A&A...276..673S}
\newcommand{\palmer}{2004AIPC..727..663P}
\newcommand{\barthelmy}{2005SSRv..120..143B}
\newcommand{\soderberg}{2009astro2010S.278S}
\newcommand{\bloom}{2009astro2010S..20B}
\newcommand{\mcquinn}{2009astro2010S.199M}

\begin{document} 
  \maketitle 

\begin{abstract}

The primary instrument of the proposed EXIST mission is a coded mask high energy telescope (the HET), that must have a wide field of view and extremely good sensitivity. In order to achieve the performance goals it will be crucial to minimize systematic errors so that even for very long  total integration times the imaging performance is close to the statistical photon limit. There is also a requirement to be able to reconstruct images on-board in near real time in order to detect and localize gamma-ray bursts, as is currently being done by  the BAT instrument on Swift. However for EXIST this must be done while the spacecraft is continuously scanning the sky. The scanning  provides all-sky coverage and is also a key part of the strategy to reduce systematic errors.  The on-board computational problem is made even more challenging for EXIST by the very large number of detector pixels (more than 10$^7$, compared with 32768 for BAT). The EXIST HET Imaging Technical Working Group has investigated and compared numerous alternative designs for the HET. The selected baseline concept meets all of the scientific requirements, while being compatible with spacecraft and launch constraints and with those imposed by the infra-red and soft X-ray telescopes that constitute the other key parts of the payload. The approach adopted depends on a unique coded mask with two spatial scales. Coarse elements in the mask are effective over the entire energy band of the instrument and are used to initially locate gamma-ray bursts. A finer mask component provides the good angular resolution needed to refine the burst position and reduces the cosmic X-ray background; it is optimized for operation at low energies and becomes transparent in the upper part of the energy band where an open fraction of 50\% is optimal. Monte Carlo simulations and analytic analysis techniques have been used to demonstrate the capabilities of the proposed design and of the two-step burst localization procedure.
\end{abstract}


\section{INTRODUCTION}
\label{sec:intro} 

The Energetic X-ray Imaging Survey Telescope (EXIST ) mission \cite{2009AIPC.1133...18G}  is optimized for study of Gamma-Ray Bursts (GRBs) as probes of the high-{\it z} Universe\cite{\bloom, \mcquinn} but will also contribute to a wide range of  other science ({\it e.g.} Refs. \citenum{2009astro2010S.105G, \soderberg, 2009astro2010S..55C})  by conducting a sky-survey in the hard X-ray /soft gamma-ray bands. 
The main instrument, the high energy telescope (HET) \cite{hongspie} 
is a coded-mask telescope operating in the band  5--600 keV and has a very wide field of view -- 1.2 sr at any instant (to 10\% effective area),  extended by continuous scanning. It is complemented by narrow field X-ray and optical/infrared instruments, the SXI covering 0.1--10 keV  \cite{tagliaferri} 
and the IRT (0.2--2.2 $\mu$m)\cite{\kutyrev}.  
  
   \begin{figure}
   \begin{center}
   \begin{tabular}{c}
 \includegraphics[trim = 0mm 0mm 0mm 0mm, clip, height=7.5cm]{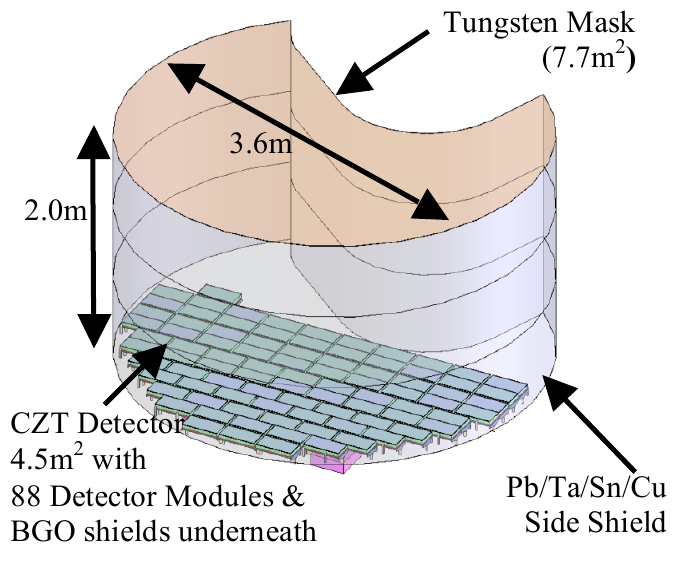}
   \end{tabular}
   \end{center}
   \vspace{-3mm}
   \caption[example] 
   { \label{fig:het_layout} 
The basic layout of the EXIST HET coded mask instrument.}
   \end{figure} 

The GRB and sky-survey aspects of the mission each present important challenges related to the associated HET image reconstruction and processing. For the GRB observations images must be reconstructed on board in very close to real time in order to detect the appearance of GRBs  or new transient  sources and to automatically initiate a slew to bring the location of the detected event within the field of view of the narrow-field instruments. The 2 year survey phase of the 5 year mission will involve combining large amounts of data together, and it is important that systematic effects are kept to an absolute  minimum to avoid them becoming a dominant source of uncertainty when the Poisson noise is reduced to a very low level.

The EXIST HET Imaging Technical Working Group, of which the authors of this paper are members, investigated and compared numerous alternative designs for the HET. We discuss here the issues considered and how image processing requirements and imaging-mode sensitivity influenced the choice of adopted design.

\section{THE HET IMAGING DESIGN and the logic behind it} 

Although numerous alternatives have been considered and assessed, the concept adopted for the HET is a very simple one -- a plane coded mask is supported 2 m above a plane detector array and parallel to it.  The mask is as large (7.7 m$^2$) as possible within the constraints of the launch vehicle shroud and other instruments. The detector array is almost as large, though limited to 4.5 m$^2$ of useful geometric area by practical considerations (required space between modules, complexity, and mass). Similarly, the detector pixel size  (0.6 mm $\times$ 0.6 mm) as small as  practical  considerations permit. 

\begin{figure}
   \begin{center}
   \begin{tabular}{c}
 \includegraphics[trim = 30mm 30mm 10mm 10mm, clip, height=6cm]{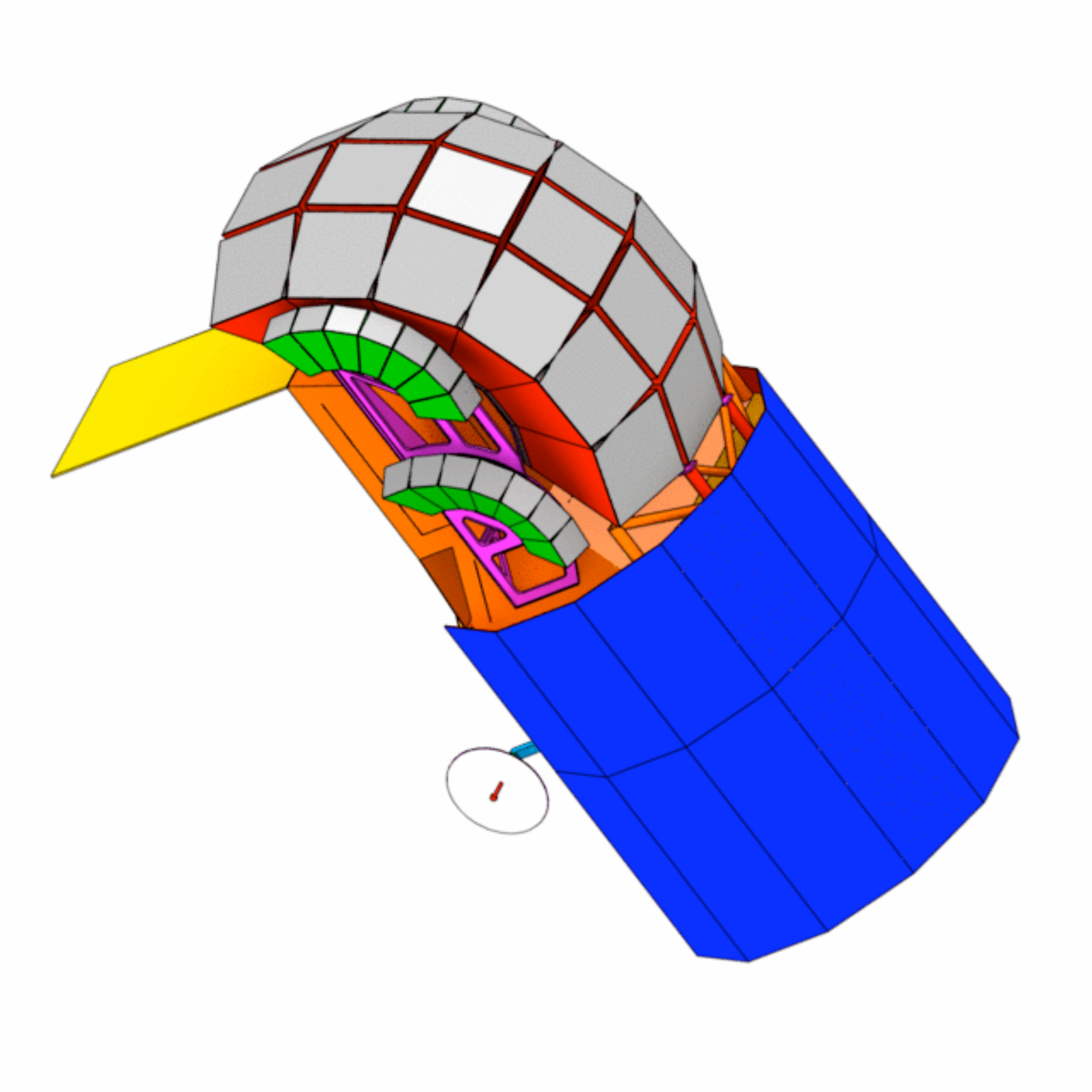}
 \includegraphics[trim = 0mm 30mm 10mm 20mm, clip,height=6cm]{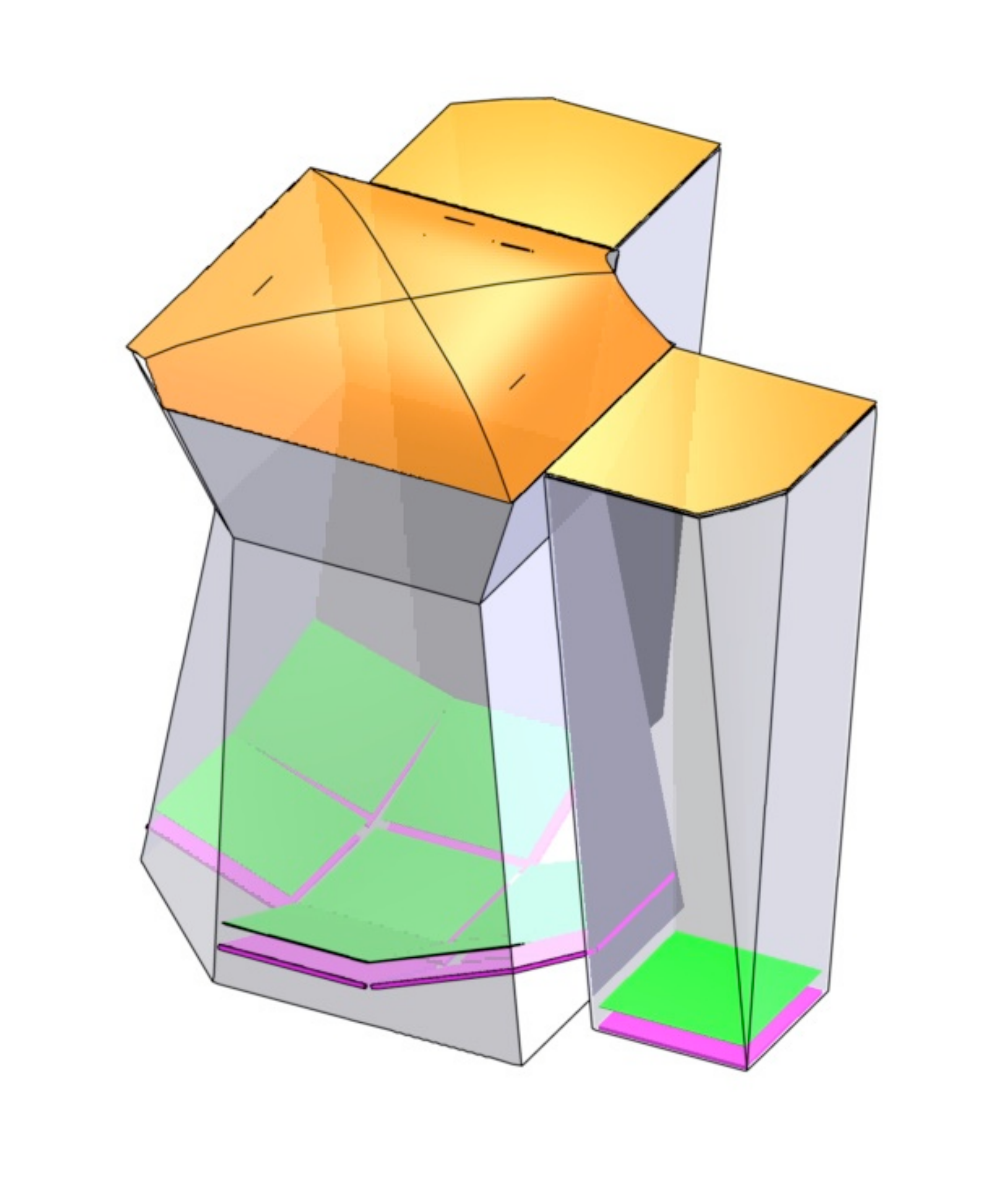}
   \end{tabular}
   \end{center}
   \caption[example] 
   { \label{fig:het-rejected} 
Two HET instrument concepts considered (a) Multiple sub-telescopes\cite{2005NewAR..49..436G} (b) A shared mask instrument (here the version considered in the 2008 Astrophysics Strategic Mission Concept Study, with two additional sub-telescopes).}
   \end{figure}

In many of the instrument concepts considered but not finally adopted, the mask and/or detector were divided  into sub-units, angled differently in an attempt to obtain an even larger field of view.  At one extreme (Fig. \ref{fig:het-rejected}a) such designs may have entirely independent subunits, with different fields of view and with shielding walls such that radiation entering through the mask of one unit cannot reach the detector of another. At the other extreme a single multi-facetted mask may cast shadows onto an array of   detector sub-units, oriented so as to view predominantly in different directions (e.g. the wide field  instrument in Fig. \ref{fig:het-rejected}b).

It was found that designs with multiple independent sub-telescopes suffered a weight penalty because of all the shielding and did not reap the full benefits of the  coded mask technique, which offers the largest multiplex advantage when the field of view is as wide as possible. 

The difficulty with designs with `shared' masks is less obvious. and is entirely associated with practicability of data analysis.  Image reconstruction from coded mask telescope data is basically a matter of correlation of the pattern recorded in the detector plane with that expected from trial source positions in each pixel in the sky image. With millions of detector pixels and millions of image pixels, this is potentially a huge problem. With a classical coded mask telescope, such as the narrow field instruments in Fig. \ref{fig:het-rejected}b, the shadow cast on the detector by sources in different directions within the field of view are simply shifts of that from an on-axis source. Thus all the correlations to be performed are those between the recorded data pattern and different shifts of a representation of the mask pattern. As will be discussed below, this is a process that can be performed very efficiently by a Fast Fourier Transform (FFT).  However, if the mask and detector are not plane and parallel, as in the main instrument in  Fig. \ref{fig:het-rejected}b, there is no such simple relationship between the shadows cast by sources in different directions. At best, FFTs can be used to give approximate solutions over a limited region of sky.

In principle, back-projection techniques\cite{\fenimore} allow the analysis of data  from coded mask systems with arbitrary geometry and they have been proposed for designs even more extreme than some of those considered for the EXIST HET (e.g. Ref. \citenum{\bird}).  However, as will be seen below, the computational requirements for its application to the HET preclude its use for  on-board imaging, though it will be see below that it can play a valuable r\^ole in refining GRB locations. Thus despite the constraints that it imposes, the  efficiency of the FFT is  crucial and it is this that drove the EXIST design in the direction of plane parallel mask and detector planes.

 Other critical parameters for the imaging design are  the mask to detector separation or  `focal length', $f$, the size, $m$,  of the mask  holes and the `open fraction' $q$ of the mask. We here consider the detector pixel size $d$ to be already fixed at a value that is as fine as is possible given available  detector technology, and cost and complexity limits. For the present purposes that is taken to be 0.6 mm.  The considerations that led to the choices made are:
 \begin{enumerate}
\item [$m$]{The dependence of the angular resolution on $m$ implies that it should be as small as possible, but the loss in sensitivity when the detector pixels imperfectly resolve the shadowsof the holes\cite{\skinnercapri}  limits how far it is sensible to go in this direction. For a source detected with adequate significance, it can be shown that the choice of $m=d$ optimizes the source location accuracy\cite{\skinnerao}, but this choice reduces the sensitivity to 2/3 of that possible with $m\gg d$ and so fewer sources will be detectable. A value of $m=1.15$ mm has been baselined, corresponding to $m/d$=1.9 and a sensitivity factor\cite{\skinnercapri}  $S/S_\infty =1-d/3m=$\ 83\% for a mask with all holes of this size. However, as will be seen below, the mask pattern adopted improves on this.} 

\item [$f$]{ Given that the extent of the mask and of the detector are limited by the shroud and by other instruments, the shorter the focal length the wider the field of view. On the other hand the angular resolution at the center of the field of view is approximately  $\sqrt{d^2 + m^2} /f$  and so becomes poorer for small $f$. The compromise adopted is $f$=200 cm.}
\item[$q$]{Equations for the optimum value of the open fraction have been given in the literature (see \citenum{\skinnerao} and references therein). Where the detector background is dominated by effects not dependent on the sky exposure ({\it e.g.} background due to high energy particles) the optimum is $q$=50\%. However where the diffuse X-ray background is important -- generally at lower energies and for wide fields of view --  a lower $q$ is preferable. Lower values of $q$ also reduce the detected event rate and hence data handling problems.  The adopted design has $q\sim0.2$ at low energy and $\sim$0.5 at high. }
\end{enumerate}

\section{IMAGING WITH A SCANNING CODED MASK TELESCOPE} 
\label{sec:twobeam}

The fact that for much of the mission EXIST will be scanning the sky has many advantages. For the detection of phenomena that are short-lived (but not instantaneous) the region surveyed is larger. For longer term observations the effects of most types of systematic noise are smeared out and drastically reduced.
However it also makes  the data analysis, particularly the on-board near-real-time analysis, more of a challenge. 

For back-projection techniques there is simply the added complication that the mapping onto the sky of the mask, as seen by the detector element in which the photon is detected, is continuously changing. For correlation image reconstruction, by FFT or otherwise, the implications of scanning are more serious. Such techniques involve binning the data and the interval over which the situation can be considered static can be very short when the angular resolution is high. A separate reconstruction must be performed for each such time bin and the resulting  tangent plane images with different centers must be combined.   
A prototype scanning coded aperture imaging analysis system has been  developed  for the Swift/BAT Slew Survey (BATSS) 
and demonstrated\cite{copete}.

\begin{figure}
   \begin{center}
   \begin{tabular}{c}
 \includegraphics[trim = 5mm 5mm 5mm 10mm, clip, height=7cm]{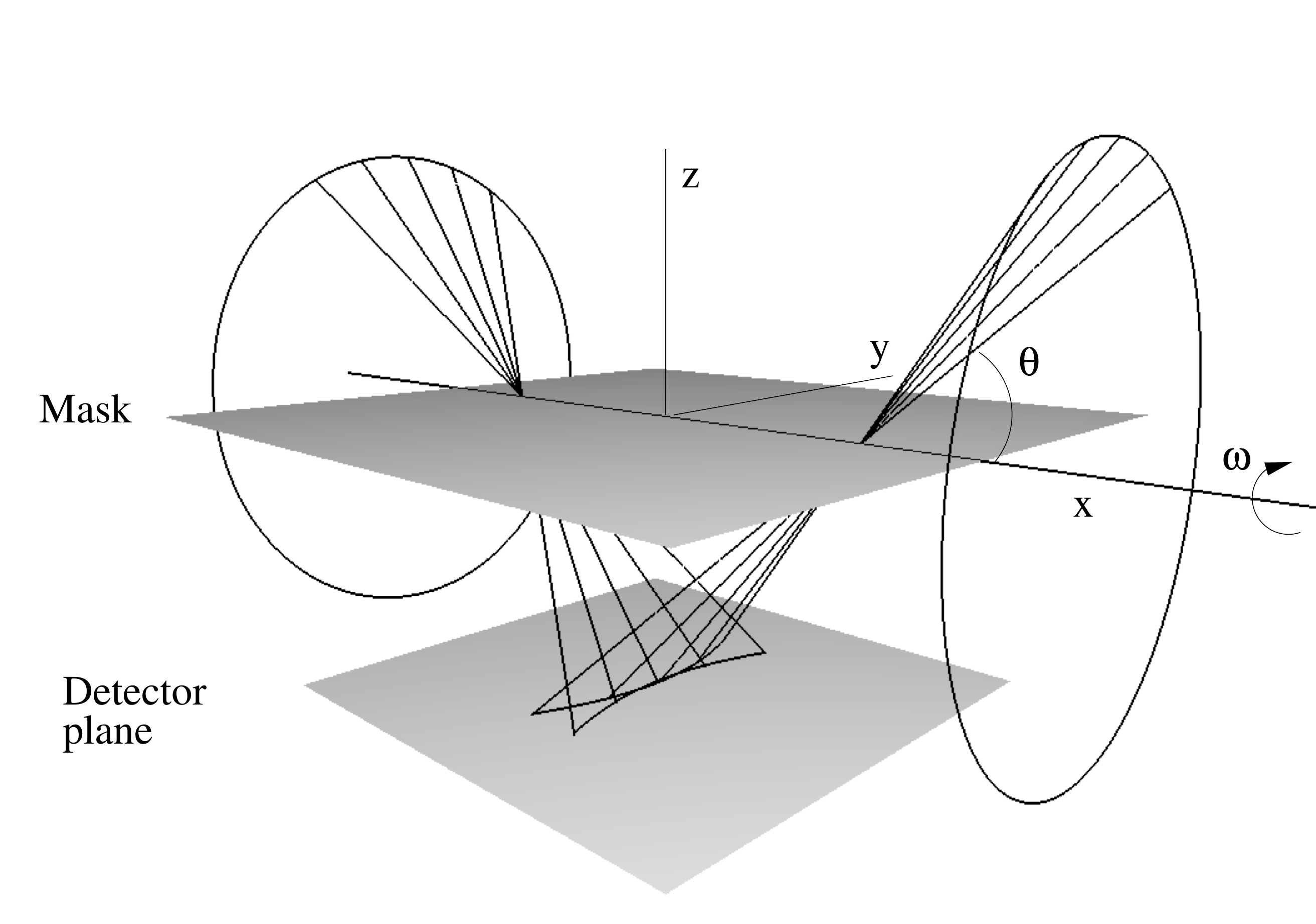}
   \end{tabular}
   \end{center}
   \caption[example] 
   { \label{fig:rotations} 
   As the instrument scans, the track across the detector of the shadow of the mask depends on the direction of the source. Here for simplicity the scanning is assumed to be  a simple rotation about an axis parallel to the mask. Sources  in the two example directions indicated will appear to move in circles on the sky and so the tracks are parabolas formed where cones intersect the detector plane. They are traversed at non-uniform speed.  }
   \end{figure}

The interval over which data may be binned can be extended somewhat using ``Time Domain Integration'' (TDI)  --  taking advantage of the fact that to a first approximation the shadows of the mask cast by sources in different directions move in similar ways. The common motion is removed before the photons are binned. Two considerations limit the scan angle over which such corrections can be made: (i) at the limits of the detector plane, or where there are gaps in the detector plane such as those between sub-units in the EXIST detector array, photons are lost ``off the edge'' (ii) the tracks if the shadows from different sources are not exactly the same (see Fig. \ref{fig:rotations}), nor are they traversed at exactly the same speed.

If the rotation is at an angular rate $\omega_x$ about an axis ($x$) normal to the instrument axis ($z$), the position in the detector of the shadow of a mask element at $(x_m,y_m)$ cast by a source at an angle $\theta$ from the rotation axis and at an azimuth angle $\phi$  about the $x$-axis at time $t=0$ is given by 
\begin{eqnarray}
x_d&=& x_m- \frac{\displaystyle{f}}{\displaystyle{\tan\theta \: Cos(\omega_x t-\phi)}}\\
y_d&= &y_m - f\:\tan(\omega_x t-\phi)
\end{eqnarray}
Thus for $t-t_0$ that is not too large,  $x_d$ is {\it approximately} constant and $y_d$ is {\it approximately} proportional to time. But sources with a range of $\theta$ and $\phi_0$ contribute to the flux received in a given part of the detector, so even after correction some blurring will remain.
Using typical numbers for EXIST shows that errors can be kept below one detector pixel for sources up to 40\degr\  from the center of the field of view over integrations up to $\sim$1 s. Without TDI the corresponding limit is 0.3 s for full resolution imaging (but see \$5).

\section{THE ON-BOARD PROCESSING PROBLEM} 
\label{sec:onboard}

The numbers of `operations' \footnote{The term `operation' here has different meanings
in different cases but typically involves several look-ups, a mathematical operation and storing the results.} for image reconstruction  using  brute force cross correlation is given by
\begin{equation}
N_{cc} \sim N_dN_s,
\label{eqn:bruteforce}
\end{equation}
where $N_d$ is the number of detector elements. For reconstruction by FFT it is
\begin{equation}
N_{FFT} \sim 2\;N_{xy} Log_2(N_{xy})
\label{eqn:fft}
\label{eqn:nlogn}
\end{equation}
where 
\begin{equation}
N_{xy} =\left[k N_d(x)+N_s(x)+B_x\right]\cdot \left[kN_d(y)+N_s(y) +B_y\right]
\end{equation}
and where the maximum dimensions of the detector and sky arrays  are $N_d(x)\times N_d(y)$ and  $N_s(x)\times N_s(y)$. Here $k$ is an oversampling factor because for the FFT to be used for the reconstruction  the detectors pixels and the sky pixels must fall on a regular grids (in space coordinates and in tangent plane respectively) whose spacings are commensurate.  Normally $k\ge2$ for grids that are sufficiently fine that sources are not missed by being split between pixels. Finally $B_x$ and $B_y$ are the widths of  border regions, padded with zeroes, introduced because the terms in square brackets must have simple factors -- indeed Eqn. \ref{eqn:nlogn} supposes that they are powers of 2.

For image reconstruction by back-projection, the number of operations depends on the number of photons detected, $N_p$ and the number of sky pixels, $N_s$ for which a reconstruction is needed:
 \begin{equation}
N_{bp} \sim N_pN_s.
\label{eqn:backproj}
\end{equation}

The numbers given by equations \ref{eqn:bruteforce} and \ref{eqn:fft} must be multiplied by the number of energy channels  and divided by the integration time to give an indication of the number of operations per second needed. In contrast, provided enough memory is available, images can be reconstructed by back projection (Eqn. \ref{eqn:backproj}) in an arbitrary number  of energy channels simultaneously and can be extracted on different timescales without penalty.

Table \ref{table:1}  gives an idea of the order of magnitude of the on-board computing task in terms of `operations' per second. Processors that are space-qualified or currently being space-qualified have been identified that should make the FFT solution feasible, but the other solutions considered would require  $\sim$10$^{3}$ to $10^6$  times more computing power. As noted above, the FFT solution demands a  plane-parallel mask and detector.

Although image reconstruction by FFT appears marginally possible using the parameters in Table \ref{table:1}, there is relatively little margin, and anyway higher rates and more energy channels are desirable. For this reason a mask design allowing a two step approach has been adopted.

 \begin{table}[htdp]
\caption{Assumptions and  the estimated size of the computing task.}

\label{table:1}
\begin{center}
\begin{tabular}{|c|c|c|c|}
\hline
Mask physical extent                   & $M_x\times M_y$     &  3380$\times$2700     & mm                         \\
 Detector physical extent             & $D_x\times D_y$      &   3380$\times$1860   & mm                        \\
Bin size                                          &   $  \delta   $                     &  0.6           &  mm \\
Corresponding angular scale\footnotemark[1]  &    $\delta/f$           &              1              & arc min  \\ 
Number of sky pixels                    &   $N_s=(M_x+D_x) (M_y+D_y)/\delta^2$     &    $8.6\times10^7$  &                                  \\
Event rate                                         &       $ C $     &     40000        &   events s$^{-1}$                     \\
Number of actual detector pixels &      $N_d$         &       1.2$\times 10^7$     &                        \\
Number of energy channels              &      $N_E$          &      4     &                        \\
Image rate                                        &    $R$           &        1     &     s$^{-1}$                  \\
Padded array size for FFT            &   $N_{xy}=N_s$, rounded up to 2$^n$           &   2$^{27}$= 134217728           &                        \\
\hline
\multicolumn{4}{|c|}{   `operations' per second   }\\
\hline
Back projection                              &    $N_{bp}=C\: N_s$            &     3.4$\times 10^{12}$          &                 \\
Convolution (no FFT)                    &   $N_{cc}=N_d\: N_s\: N_E\: R$              &   4.1$\times 10^{15}$            &      operations s$^{-1}$                \\
Convolution with FFT                    &   $N_{FFT}=2\: N_{xy}\: Log_2N_{xy}$               &  7.2$\times 10^9$            &                       \\
\hline

\end{tabular}
\\
\vspace{2mm}
\footnotemark[1]{\footnotesize at center of FOV; reduced in proportion to $(1+\sec^2\theta)^{-1}$ at off-axis angle $\theta$.}
\end{center}

\end{table}%

\section{THE HYBRID MASK AND TWO-STAGE SOURCE LOCATION} 
\label{sec:two-stage}

\begin{figure}
   \begin{center}
   \begin{tabular}{c}
 \includegraphics[trim = 16mm 16mm 16mm 10mm, clip, height=10cm]{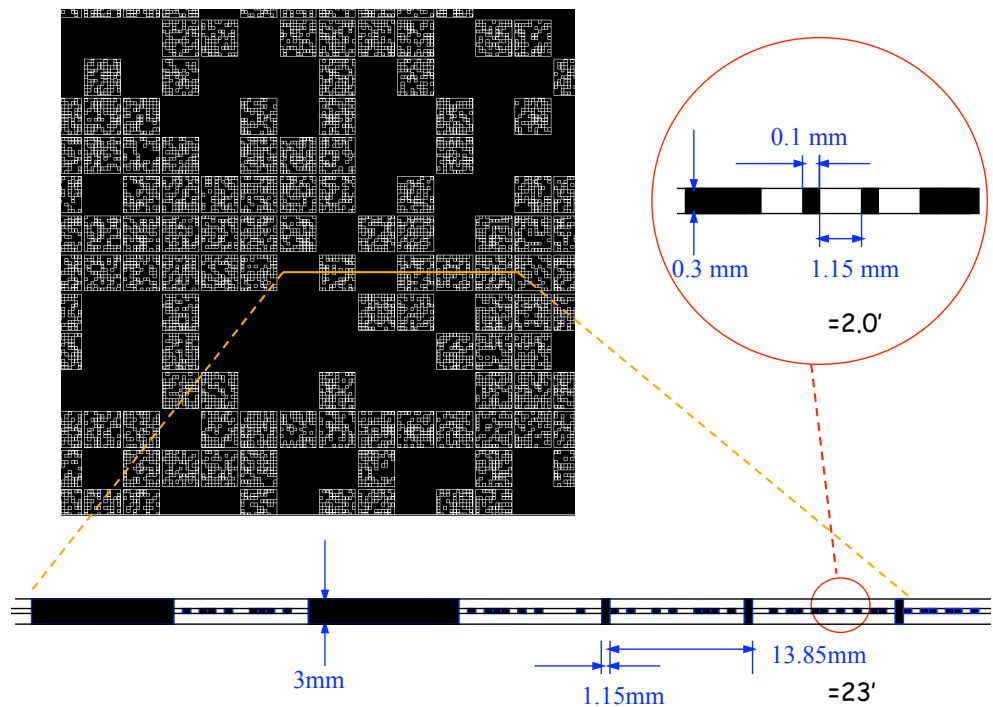}
   \end{tabular}
   \end{center}
   \caption[example] 
   { \label{fig:twoper} 
   A small section of the hybrid mask proposed for the EXIST HET.   }
   \end{figure}

The  concept baselined for the EXIST HET  involves  a hybrid mask containing patterns on two spatial scales\cite{\skinnergrindlay} illustrated in Fig. \ref{fig:twoper}.  It has a  large-scale (coarse) structure with a random pattern of elements on a 15 mm pitch, half of which are solid 3 mm thick tungsten. The `open' elements of the coarse pattern  contain thinner (0.3 mm) tungsten masks with random holes on a 1.25 mm pitch. Again 50\% of the elements are solid\footnote{All of these dimensions and parameters are illustrative and may be subject to further optimization.}. The thickness  of the coarse mask elements is chosen so that they are  as opaque as possible over the entire operating energy range, while staying within  mass constraints. The thin mask sections are designed to become largely transparent in the upper part of the energy range.
The objectives of this design are to
\begin{enumerate}
\item { ensure that both at low energies and at high the `open fraction' of the mask is close to optimal (\S\ref{sect:fopt})}
\item { greatly reduce the computing power needed using a two step processing strategy   (\S\ref{sect:twostep})}.\\
\end{enumerate}
\vspace {-5mm}
{\noindent {In addition, it}}
\begin{enumerate}
\setcounter{enumi}{2}
\item{largely  avoids an `auto-collimation'  problem -- if holes as small as those in the fine mask sections (1.15 mm) were to be made in material thick enough to be largely opaque at high energies  the field of view would be severely limited }
\item{ reduces the number of events that have to be handled by the data processing and telemetry by having a low open fraction at low energies where the Cosmic X-ray background (CXB)  dominates the counting rate,  without loss of sensitivity except for very high time resolution studies of extremely bright sources.}
\end{enumerate}

\subsection{Optimizing the open fraction}
\label{sect:fopt}

\begin{figure} [tb]
   \begin{center}
   \begin{tabular}{c}
 \includegraphics[trim = 0mm 10mm 0mm 0mm, clip, width=12cm]{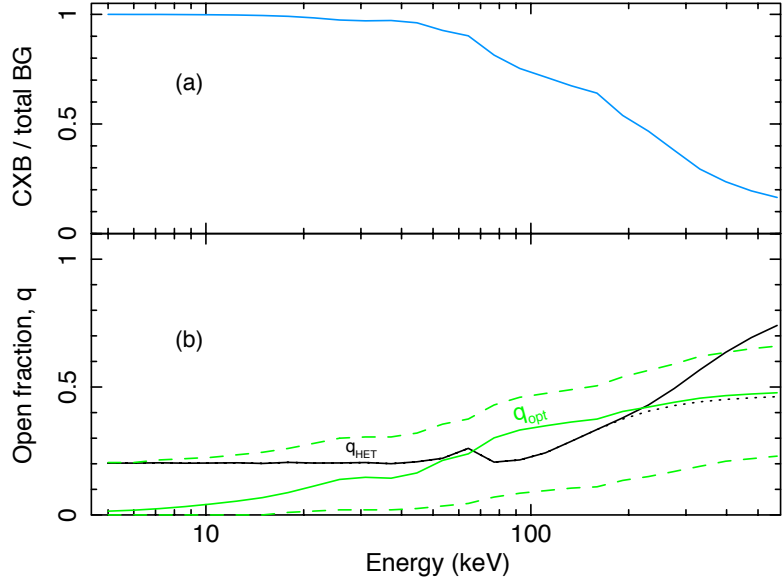}
   \end{tabular}
   \end{center}
   \caption[example] 
   { \label{fig:open_frac} (a) The fraction of the total  expected  background in the HET due to aperture-dependent effects (e.g. CXB). The values given are for a fully open aperture (no mask, i.e. $q$=1). (b) The optimum value of the open fraction, $q$ as a function of energy, based on the background estimates in (a).  It is assumed that the noise is dominated by background, not by source counting statistics, which will generally be true for the HET.  The dashed lines indicate the band within which the sensitivity is within 10\% of that for the optimum $q$.  Also shown is the transmission (on-axis) for the mask design in Fig. \ref{fig:twoper}. The solid line is if leakage through the thick coarse mask elements is ignored; the dotted line shows the effect of including it.  }
   \end{figure}

Fig. \ref{fig:open_frac}a shows that at energies below about 100 keV  the detector background due to aperture-dependent effects (direct and indirect effects of the CXB) is expected to dominate over  non-aperture-dependent background (particle-induced events, activation and indirect effects of penetrating high-energy gamma-rays). The fraction due to the former is shown as if there were no mask and in practice will be lower by a factor $\sim$$q$.  In these circumstances $q<$50\% offers a sensitivity advantage, though in practice never a very large one \cite{\skinnerao}. Fig. \ref{fig:open_frac}b shows the optimum $q$ as a function of energy, together with the on-axis  transmission $t$ of the hybrid mask. At the lowest energies $t>q_{opt}$ because it becomes constant at $\sim$25\% (actually a little lower due to the supporting bars between elements). However it is always such as to give a sensitivity within 10\% of that for $q_{opt}$. In fact the calculation of $q_{opt}$ assumes that one is not limited by the number of source photons; for studying bright short bursts and QPOs there is an advantage in avoiding extremely low values of $q$. At the top end of the energy band it is again the case that  $t>q_{opt}$. This arises because of leakage though the 3 mm thick coarse mask elements. The associated loss in sensitivity is inevitable given that mass limitations preclude making the  elements thicker.

\subsection{Imaging and source location}
\label{sect:twostep}

The most challenging  objective of the HET on-board processing is the detection of GRBs and their localization as quickly as possible so that the satellite can be slewed to bring them within the fields of view of the other instruments. The principle and approach are very similar to those of Swift/BAT\cite{\barthelmy, \palmer}, but the size of the computing task is considerably greater. HET will have 1.2$\times$10$^7$ detector pixels, compared to BAT's 32768 and the  number of operations per FFT, $N_{FFT}$, is nearly 200 times larger. In addition, the objective is to obtain $\sim$1 image per second wheras BAT requires 6 sec to form an image and on short timescales can only detect GRBs when a trigger has been first detected by monitoring event rates. 

   \begin{figure} [tb]
   \begin{center}
   \begin{tabular}{c}
(a)  \includegraphics[trim = 0mm 20mm 0mm 0mm, clip, width=6cm]{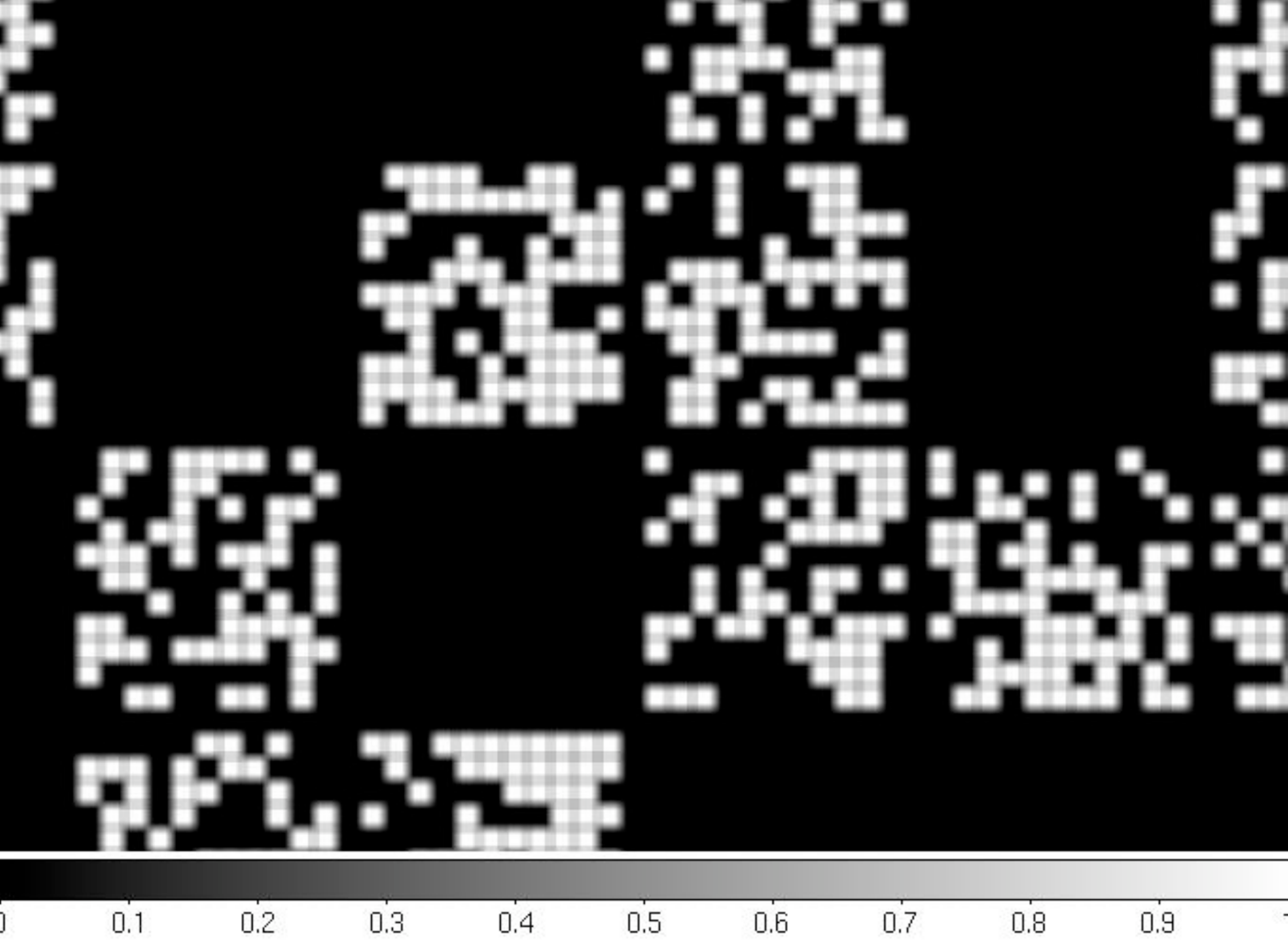}
 \includegraphics[trim = 0mm 20mm 0mm 0mm, clip, width=6cm]{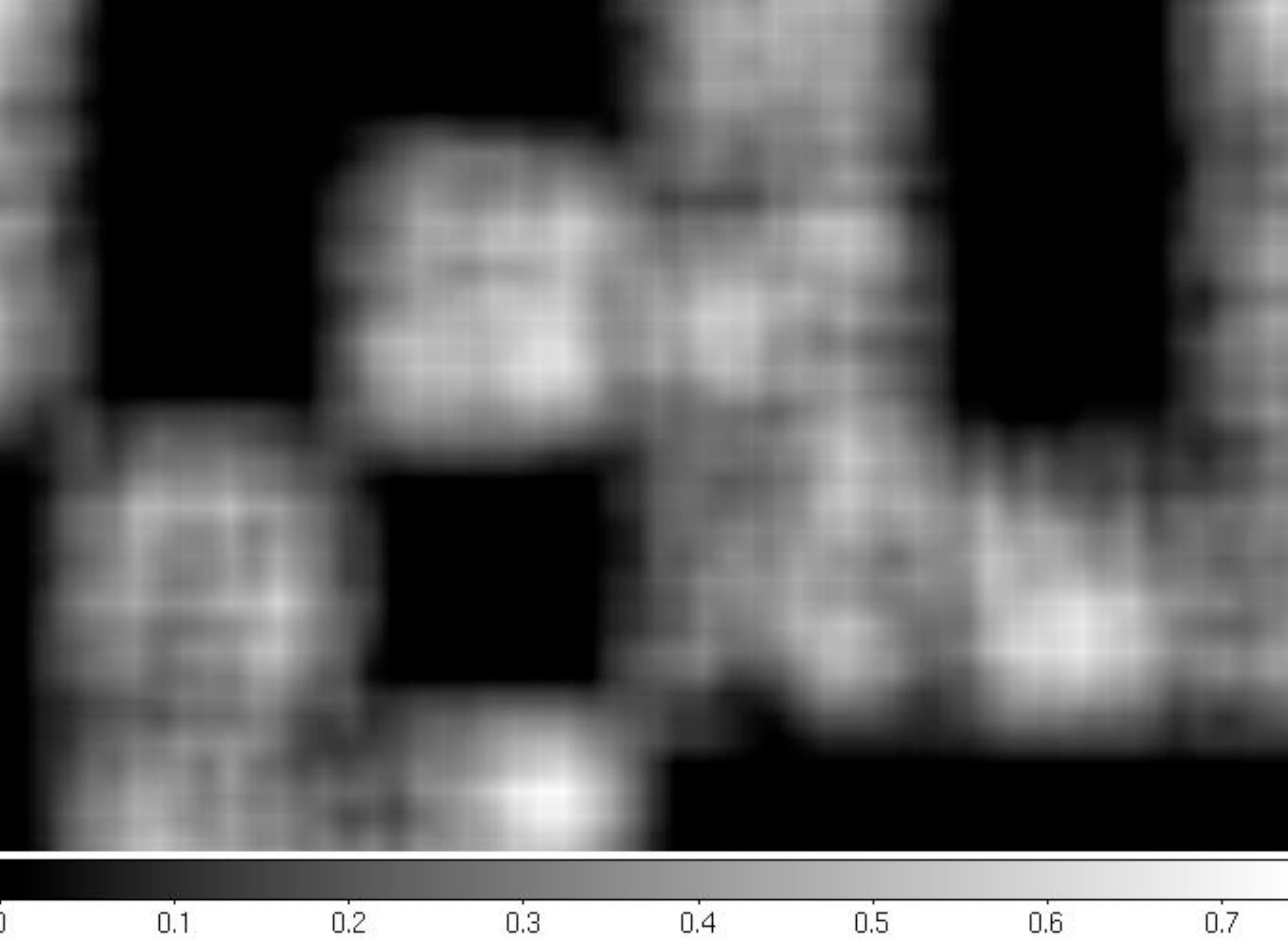}  (b)\\
(c)  \includegraphics[trim = 0mm 20mm 0mm 0mm, clip, width=6cm]{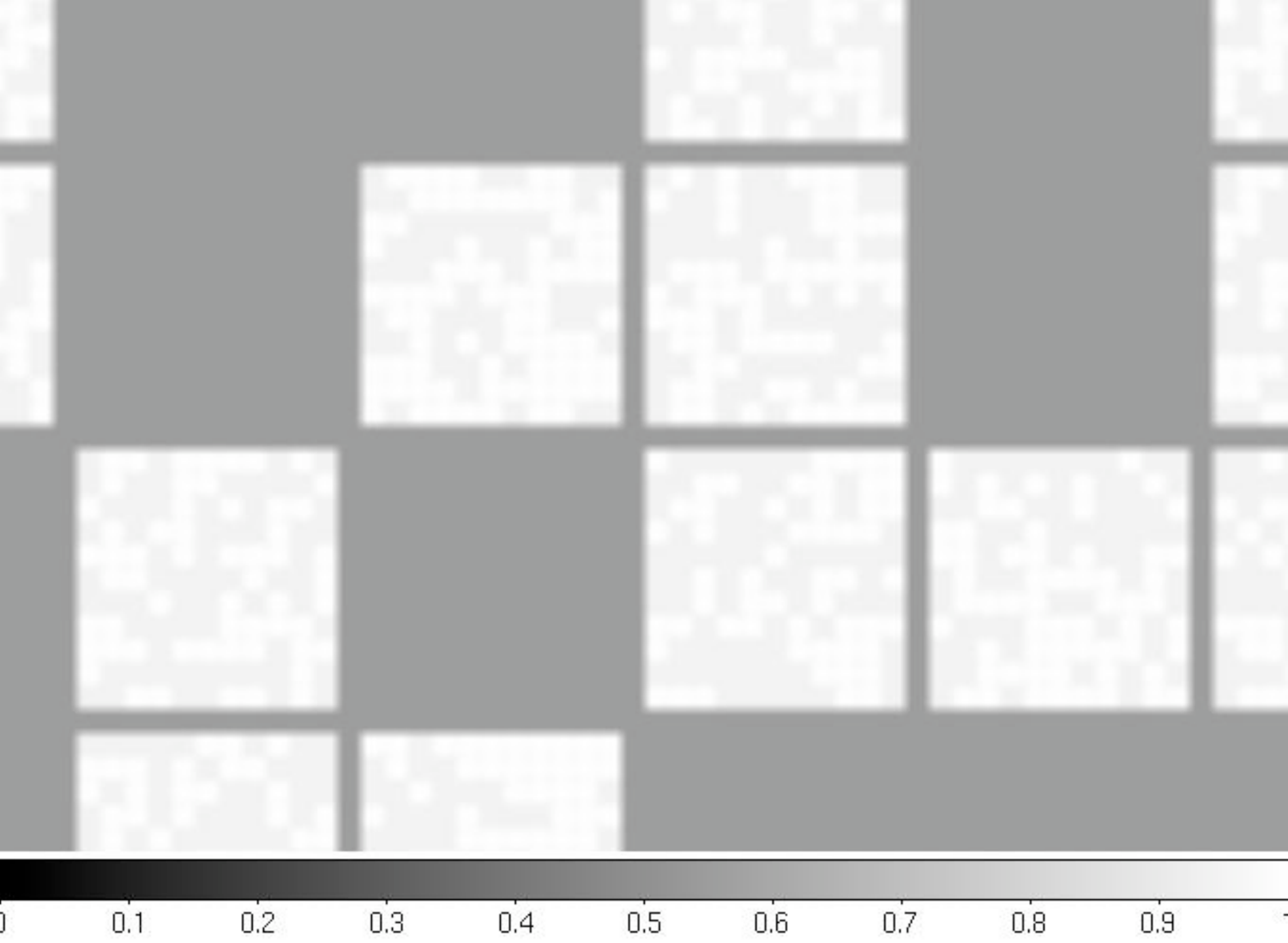}
 \includegraphics[trim = 0mm 20mm 0mm 0mm, clip, width=6cm]{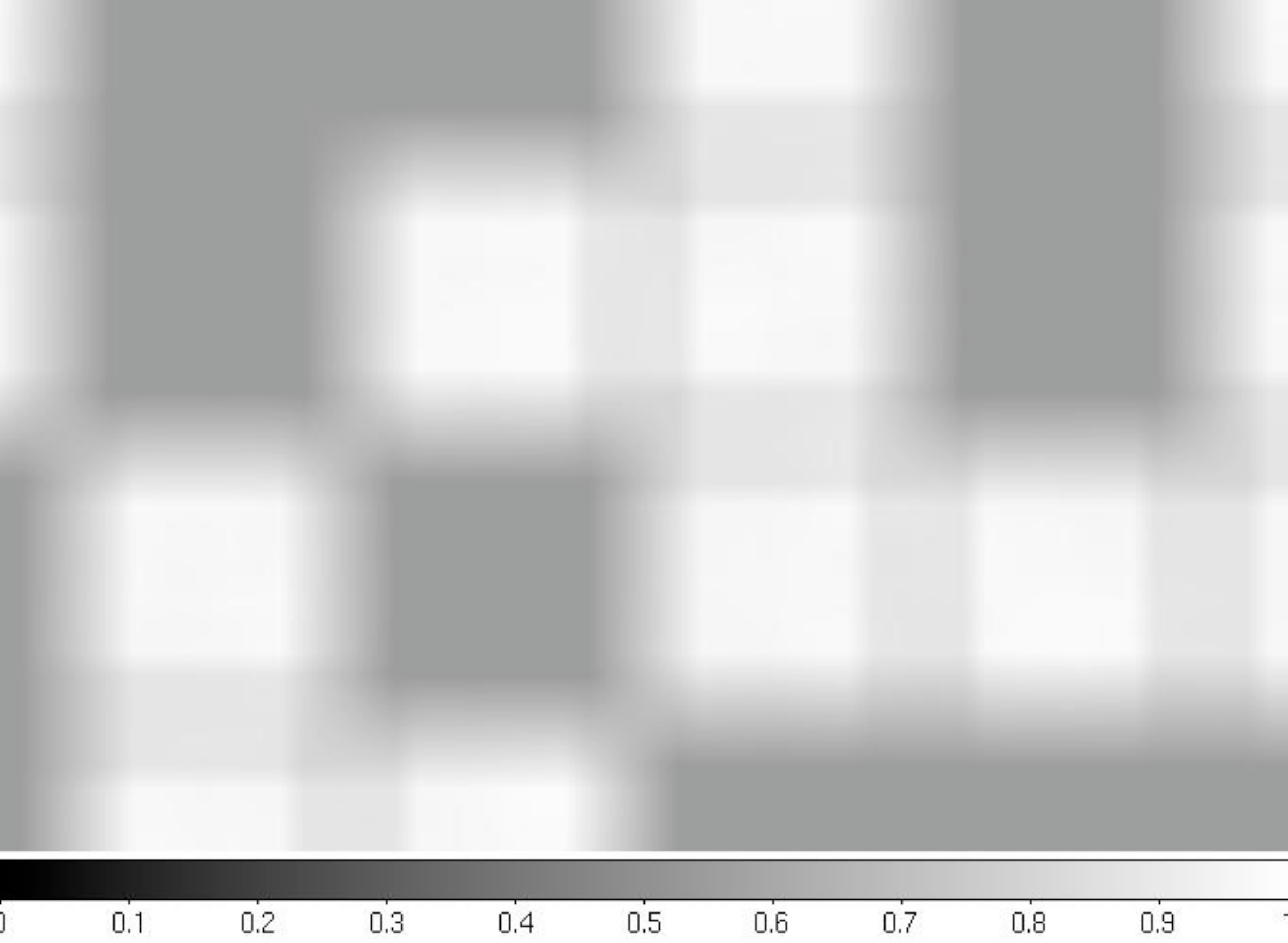}(d)\\
  \includegraphics[trim = 5mm 0mm 5mm 160mm, clip, width=12.2cm]{Figs/high_energy_6mm_dcf_msk.pdf}
   \end{tabular}
   \end{center}
   \caption[example] 
   { \label{fig:dcf_msk} Part of the hybrid mask, seen at different energies and by a detectors with different  spatial resolution. Top row (a,b): low energy (5keV). Bottom row (c,d): high energy (600 keV). Left (a,c): 0.6 mm detector resolution. Right (b,d): 6 mm resolution. The figures are convolutions of the shadow with a detector pixel; the actual response to the shadow cast by a point source would be this function, sampled on 0.6 or 6 mm grid. }
   \end{figure} 

As well as providing near-optimal open fraction at all energies, the hybrid mask provides the key to reducing the  burst detection processing problem to a manageable size.  At low energies, even the 0.3 mm thin mask sections are effectively opaque. Viewed with the full resolution of the HET detector array, it therefore casts a shadow of the form in Fig. \ref{fig:dcf_msk}a. For high energy photons,  the thin mask sections are largely transparent and only the coarse structure is seen (Fig. \ref{fig:dcf_msk}c). If the shadow is recorded with a coarsely binned detector -- the fine mask regions become `gray' and the shadow is predominantly that of the coarse mask, both at low energies  and at high (Fig. \ref{fig:dcf_msk}b,d). 

\begin{figure} [tb]
  \begin{center}
   \begin{tabular}{c}
 \includegraphics[trim = 0mm 10mm 0mm 28mm, clip, width=12cm]{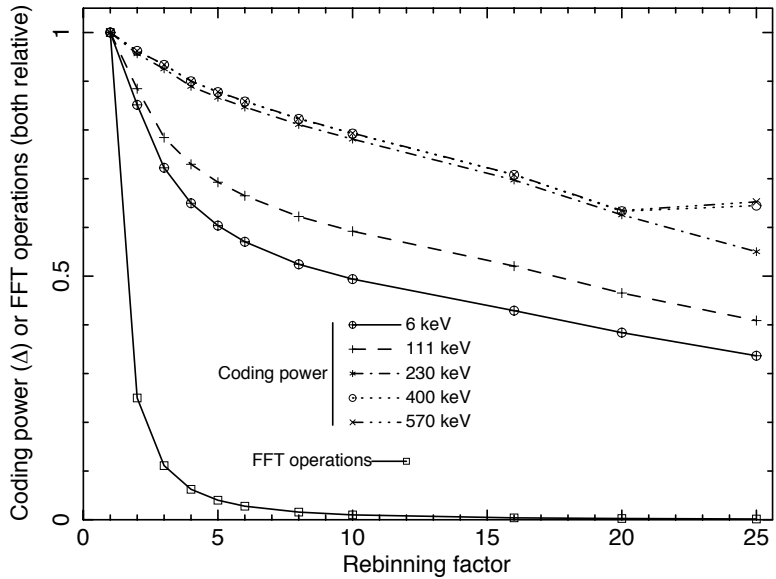}
   \end{tabular}
   \end{center}
    \caption{ \label{fig:coarse_sensitivity} The effect of using coarsely binned data for image reduction. Coarse binning reduces the coding power (and hence sensitivity) . The reduction is less at high energies, where the effective mask is predominantly the coarse one, than at low energy. The size of the computing task associated with an image reconstruction by FFT (square symbols, solid line) drops very rapidly.   }
   \end{figure}

Consider what happens if an image is reconstructed using the coarsely binned detector data and an appropriately coarse representation of the mask pattern. The angular resolution will be approximately that corresponding to the coarse mask ($\sim$25$'$), but there will be some loss of sensitivity because not all of the information potentially available is being used. The effect on the sensitivity can be quantified using the `coding-power'. The coding power,  $\Delta$, of a mask-detector combination provides a measure of its sensitivity relative to a similar  idealized configuration and in which  the detector has the same background rate but with perfect  spatial resolution and in which the mask has a pattern that is 50\% perfectly open, 50\% perfectly opaque\footnote{It is assumed that observation is limited by background and not by source flux or counting statistics (see Ref. \citenum{\skinnerao}).}. $\Delta$ is simply equal to twice  the {\it rms} of mask transmission, as seen with the  finite resolution  detector (a number in the range 0--1). Fig. \ref{fig:coarse_sensitivity} shows how the sensitivity  decreases as detector pixels are binned together, but also how the computing load decreases much faster.

The strategy proposed for detecting GRBs in the imaging mode  involves, for each time and energy range combination, the following steps:
\begin{enumerate}
\item {If the time range exceeds that for which TDI is possible, divide the data into sub-periods.}
\item{For each sub-period:}
   \begin{enumerate}
   \item{ Bin the data into coarse detector bins, correcting for mean shadow motion (ie TDI).}
   \item{ Predict the response for each known bright source in the field of view; fit for the intensity of the source; subtract from the binned data array. }
   \item{ Reconstruct a coarse image by FFT. }
   \end{enumerate}
 \item { If there are multiple sub-periods  overlay and combine the images on various timescales ({\it e.g.} 3, 10, 30 s ...)}
 \item {Search the image for possibly significant points, using a low threshold ( $N_{\sigma 1}$).}
\item{ For each possibly significant point, make  full resolution  local images around the location by back projection, using a detailed, energy-dependent,  mask model and the actual spacecraft attitude at the time of arrival of the photon.} 
\item{Optionally subtract a reference image.} 
\item {Any  peak greater than a higher threshold ($N_{\sigma 2}$) in one of these local images is considered a valid trigger.}
\end{enumerate}

The same procedure can be used  for locating possible GRBs identified by monitoring event rates. In this case the energy band and time range will be those in which the rate increase was seen and the size of the field of view imaged initially in the coarse mask may be somewhat reduced if the candidate event was detected in a subset of the detector plane (e.g. a particular quadrant).  
 
The strategy depends on the fact that although the coarse imaging is intrinsically less sensitive, by allowing a relatively large number of false triggers, a much lower threshold can be set. For a one-step analysis the threshold $N_{\sigma 1} = 6.9$ in Table \ref{table:twostep} is chosen so that, taking into account the very large number of resolution elements in each image and the fact that one might be examining one image per second, the number of false triggers to which the spacecraft would slew is less than one per 2 days. On the other hand for the coarse imaging stage of a two-step analysis by allowing, say, 10 false triggers per coarse image, each of which will be followed-up by local fine-imaging using back-projection, the threshold can be reduced by nearly a factor of two -- to $N_{\sigma 2}=3.8$ for the figures in the table.
The loss in sensitivity associated with coarse imaging  is energy dependent (Fig. \ref{fig:coarse_sensitivity}) but except at the lowest energies and the coarsest binning is less than the amount by which the threshold can be reduced. Thus any event detectable by full fine FFT imaging is almost certain to be detected by the two-step process  with, in this example, about 40 times less computing load.

In practice a sort of two-step process could be adopted even with the fine FFT approach in that the location of a GRB would probably be found, and its significance confirmed, by follow-up  back-projection in a small region around the peak. The threshold at the FFT stage could then be reduced somewhat and the higher one applied only after the follow-up. But this does not change any of the numbers in the Table except to increase slightly the already huge  computational load associated with such an approach.

\begin{table}[tdp]
\caption{\label{table:twostep} Comparison of one-step and two-step strategies for detecting GRBs and transient sources. It is assumed that for the coarse imaging $10\times10$ detector pixels  binned  together and that typically 10 candidate peaks per coarse image are followed-up using detailed back-projection. Calculations are based on a single energy bin and one image per second. }
\begin{center}
\begin{tabular}{|c|c|c|c|c|}
\hline
                                               &  ~  ~   One--step   ~  ~                         &  \multicolumn{2}{c|}{Two-step}                                                &           Units                                \\
                                               \cline{2-4}
                                               &     Fine FFT                              &  ~ Coarse FFT            ~                     &  Fine back-projection   &                                          \\
  \hline 
\multirow{2}{*}{Image size}&  $ 5\times10^8$               &      $  5\times10^6 $              &             $1600 $             &           pixels                     \\
                                                &       $ 2\times10^7$          &          $ 2\times10^6$             &               $100$               &   resolution elements    \\
                                                \hline
Computation size               & $2\times10^{10}$                     &   $2\times10^8$                           &   $1.5\times10^8$            &           `operations'         \\      
\hline
\multirow{2}{*}{Threshold}&          7.2                                    &                    $N_{\sigma 1}=3.9$     &   $N_{\sigma 2} = 7.2 $   &   significance,  $\sigma$        \\ 
                                                &                1.0                              &         1.04$^a$ -- 0.83$^b$             &             1.0                           &   flux (relative)             \\
\hline
\multirow{3}{*}{False trigger rate$^c$} &                 $ 3\times10^{-13} $     &     $6\times10^{-5}  $         &    $3\times10^{-13} $   &    per resolution element      \\
                                                           &   $6\times10^{-6} $     &     $10 $                  &    $6\times10^{-6} $   &    per image          \\
                                                             &    0.5                                            &              ($8\times 10^5$)                     &          0.5                             &            per day                 \\
\hline
\end{tabular}\\
$^a$ low energy ~ ~ ~ $^b$ high energy ~ ~ ~ $^c$ at given threshold.
\end{center}
\label{default}
\end{table}%

\subsection{The Point Spread Function and source location accuracy}

\begin{figure}[tb]
   \begin{center}
   \begin{tabular}{c}
 \includegraphics[trim = 5mm 5mm 5mm 5mm, clip, height=8cm]{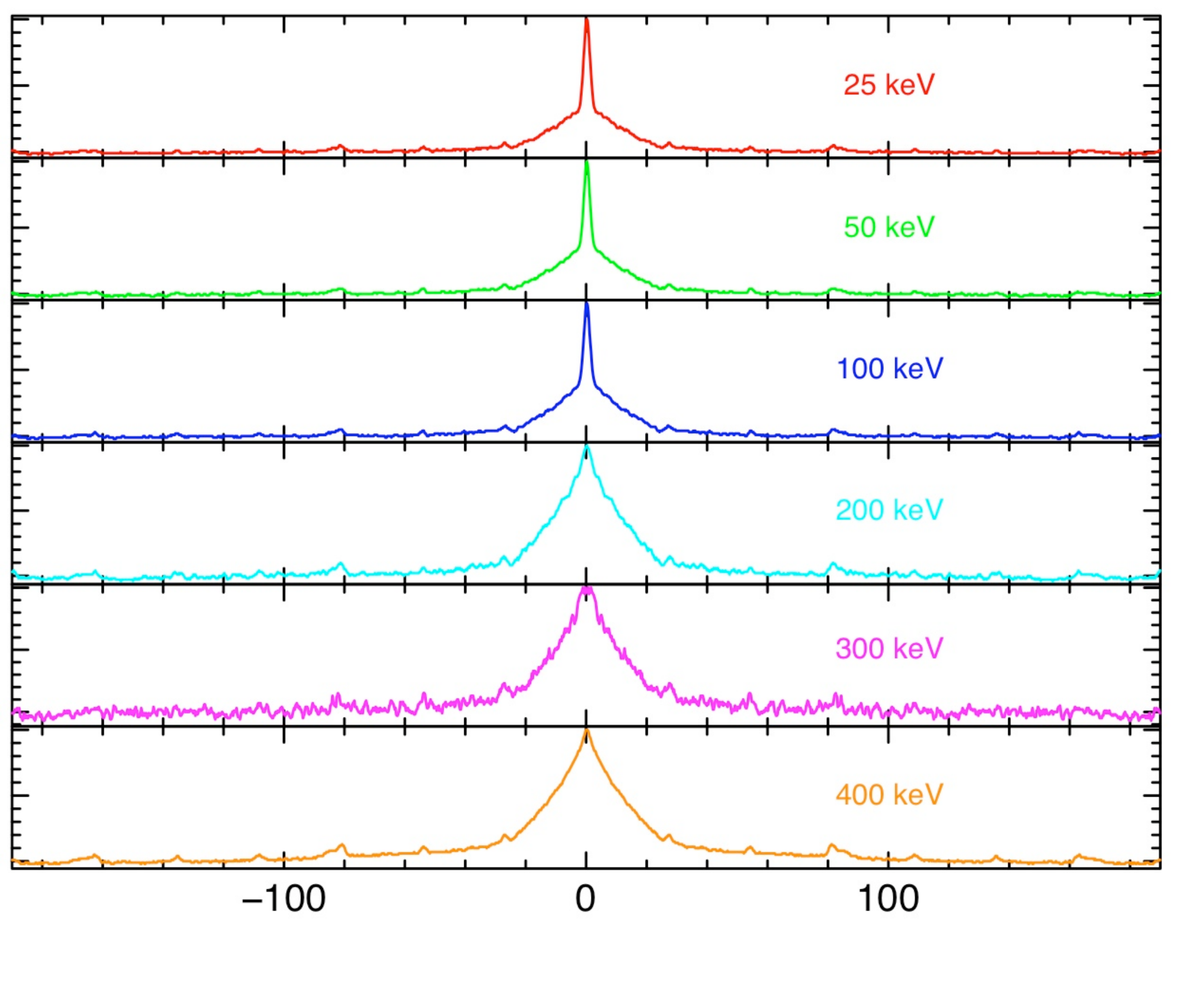}
   \end{tabular}
   \end{center}
   \caption[example] 
   { \label{fig:psfs} 
  The PSF of HET with the hybrid mask, as a function of energy. The units are arc minutes. }
   \end{figure}

The Point Spread Function  (PSF) is the normalized auto-correlation of the mask transmission convolved with the pixel footprint.   With the hybrid mask the PSF is energy-dependent, as seen from the results of Monte Carlo simulations in Fig. \ref{fig:psfs}. Together with the significance of detection, the PSF width sets a limit on the precision with which a source can be located. The 1$\sigma$ source location accuracy for a source detected at significance $N_\sigma$  is approximately the width of this function where it falls to a fraction $(1 -1/N_{\sigma})$ of the maximum. If $N_{\sigma}$ is sufficiently large that a source is detected with confidence, then in practice this means that, except for the highest energies, it is the sharpness of the central spike that matters. That sharpness is inevitably inferior to that for a similar instrument with a (computationally impracticable)  `fine-only' mask, though the loss is alleviated by the fact that the detection will be at a slightly higher significance because the finite detector resolution has less impact.
 
\section{SYSTEMATIC NOISE AND THE SKY SURVEY} 

The all-sky survey aspect of the EXIST objectives presents a different set of challenges. Data analysis will be in retrospect and on the ground, so computer power is much less of a constraint. The most important problem facing the mission design is then the minimization of systematic errors  in order to obtain a sensitivity close to the theoretical limit when data from 2 years of observations are combined together. Systematic noise occurs in images reconstructed from coded-mask telescope data in two ways (a) imperfect  modeling of the response to bright sources and hence residues left after their subtraction and (b) imperfect modeling of non-uniformities in the background. Spurious structure on a particular spatial scale, $l$,  in the array representing the data in the  detector plane leads to structure on the corresponding angular scale $l/f$ in a reconstructed image. Thus when searching for point sources it is small scale residuals that are most important.

There is now considerable experience with deep hard X-ray surveys with the coded mask instruments INTEGRAL/IBIS and Swift/BAT. Of the two, Swift/BAT  provides the closest analogy with EXIST/HET, though it has only $\sim1/10$ the detector area and only 32000 detector pixels instead of more than $10^7$.  BAT observations are divided into `snapshots' of $\sim$1000 s, with the same stable spacecraft attitude, after which there is a rapid slew to another attitude, though there may be a return to the  first attitude, frequently  one orbit later. Experience with Swift/BAT shows that near-Poisson-noise limited sensitivity can be achieved in single snapshots, but that extreme care is necessary when combining together images from  multiple  snapshots, particularly when the spacecraft attitudes are similar. 

EXIST will have two major advantages over INTEGRAL and Swift in the battle against systematic noise. First, the much greater number of detector pixels helps. To the extent that the contributions from different detector pixels  to the systematic noise are uncorrelated in sky coordinates, there is an advantage that goes with the square-root of the number of pixels -- a factor of $\sim$20. Secondly the EXIST HET will take an effectively different snapshot every $\sim$0.5 s, compared with  $\sim$1000 s   for BAT. Again assuming that the noise in sky coordinates is uncorrelated, this implies a further reduction in systematic noise by a factor $>$40.   
Another way of looking at the problem is to note that during a 2 year survey,  as a result of the scanning, each of the $1.2\times10^7$  detector pixel will acquire $~\sim10^9$ measurements in each energy band. 

\section{CONCLUSIONS}

EXIST is an ambitious mission that will detect GRBs more distant than any studied to date and up to an order of magnitude fainter. The challenges that it presents in terms of on-board detection and location of transient events and of the reduction of systematic noise can be overcome only because of the adoption of a simple geometry that makes correlation efficient and a  mask design that permits a two-step detection  and location of new sources.

\acknowledgments     
 
The EXIST mission studies involve a very large number of people. The authors are members of the EXIST HET Imaging Technical Working Group. We gratefully acknowledge the efforts of the other contributors to this large community effort.  Partial support for this study was derived from NASA-ASMC 
 grant NNG04GK33G.


\bibliography{spie_09_exist}  
\bibliographystyle{spiebib}   

\end{document}